\font\tenfrakturb=eufb10
\font\tenfraktur=eufm10
\font\tenmsbm=msbm10
\font\sevenfrakturb=eufb7
\font\sevenfraktur=eufm7
\font\sevenmsbm=msbm7
\font\fivefrakturb=eufb5
\font\fivefraktur=eufm5
\font\fivemsbm=msbm5
\newfam\bgothicfam
\newfam\gothicfam
\newfam\msbmfam
\textfont\bgothicfam = \tenfrakturb \scriptfont\bgothicfam=\sevenfrakturb
\scriptscriptfont\bgothicfam=\fivefrakturb
\textfont\gothicfam = \tenfraktur \scriptfont\gothicfam=\sevenfraktur
\scriptscriptfont\gothicfam=\fivefraktur
\textfont\msbmfam = \tenmsbm \scriptfont\msbmfam=\sevenmsbm
\scriptscriptfont\msbmfam=\fivemsbm

\def\Bbb{\tenmsbm\fam\msbmfam}

\catcode`@=11
\def\renewcounter#1{\@definecounter{#1}\@ifnextchar[{\@newctr{#1}}{}}

\documentstyle[twoside]{article}

\catcode`\@=11
\long\def\@makefntext#1{
\protect\noindent \hbox to 3.2pt {\hskip-.9pt
$^{{\eightrm\@thefnmark}}$\hfil}#1\hfill}		

\def\@makefnmark{\hbox to 0pt{$^{\@thefnmark}$\hss}}	

\def\ps@myheadings{\let\@mkboth\@gobbletwo
\def\@oddhead{\hbox{}
\rightmark\hfil\eightrm\thepage}
\def\@oddfoot{}\def\@evenhead{\eightrm\thepage\hfil
\leftmark\hbox{}}\def\@evenfoot{}
\def\sectionmark##1{}\def\subsectionmark##1{}}

\oddsidemargin=\evensidemargin
\newcounter{sectionc}\newcounter{subsectionc}\newcounter{subsubsectionc}
\renewcommand{\section}[1] {\vspace{12pt}\addtocounter{sectionc}{1}
\setcounter{subsectionc}{0}\setcounter{subsubsectionc}{0}\noindent
	{\tenbf\thesectionc. #1}\par\vspace{5pt}}
\renewcommand{\subsection}[1] {\vspace{12pt}\addtocounter{subsectionc}{1}
	\setcounter{subsubsectionc}{0}\noindent
	{\bf\thesectionc.\thesubsectionc. {\kern1pt \bfit #1}}\par\vspace{5pt}}
\renewcommand{\subsubsection}[1] {\vspace{12pt}\addtocounter{subsubsectionc}{1}
	\noindent{\tenrm\thesectionc.\thesubsectionc.\thesubsubsectionc.
	{\kern1pt \tenit #1}}\par\vspace{5pt}}
\newcommand{\nonumsection}[1] {\vspace{12pt}\noindent{\tenbf #1}
	\par\vspace{5pt}}

\newcounter{appendixc}
\newcounter{subappendixc}[appendixc]
\newcounter{subsubappendixc}[subappendixc]
\renewcommand{\thesubappendixc}{\Alph{appendixc}.\arabic{subappendixc}}
\renewcommand{\thesubsubappendixc}
	{\Alph{appendixc}.\arabic{subappendixc}.\arabic{subsubappendixc}}

\renewcommand{\appendix}[1] {\vspace{12pt}
        \refstepcounter{appendixc}
        \setcounter{figure}{0}
        \setcounter{table}{0}
        \setcounter{lemma}{0}
        \setcounter{theorem}{0}
        \setcounter{corollary}{0}
        \setcounter{definition}{0}
        \setcounter{equation}{0}
        \renewcommand{\thefigure}{\Alph{appendixc}.\arabic{figure}}
        \renewcommand{\thetable}{\Alph{appendixc}.\arabic{table}}
        \renewcommand{\theappendixc}{\Alph{appendixc}}
        \renewcommand{\thelemma}{\Alph{appendixc}.\arabic{lemma}}
        \renewcommand{\thetheorem}{\Alph{appendixc}.\arabic{theorem}}
        \renewcommand{\thedefinition}{\Alph{appendixc}.\arabic{definition}}
        \renewcommand{\thecorollary}{\Alph{appendixc}.\arabic{corollary}}
        \renewcommand{\theequation}{\Alph{appendixc}.\arabic{equation}}
        \noindent{\tenbf Appendix \theappendixc #1}\par\vspace{5pt}}
\newcommand{\subappendix}[1] {\vspace{12pt}
        \refstepcounter{subappendixc}
        \noindent{\bf Appendix \thesubappendixc. {\kern1pt \bfit #1}}
	\par\vspace{5pt}}
\newcommand{\subsubappendix}[1] {\vspace{12pt}
        \refstepcounter{subsubappendixc}
        \noindent{\rm Appendix \thesubsubappendixc. {\kern1pt \tenit #1}}
	\par\vspace{5pt}}

\newcommand{\textlineskip}{\baselineskip=13pt}
\newcommand{\smalllineskip}{\baselineskip=10pt}

\def\eightcirc{
\begin{picture}(0,0)
\put(4.4,1.8){\circle{6.5}}
\end{picture}}
\def\eightcopyright{\eightcirc\kern2.7pt\hbox{\eightrm c}}

\newcommand{\copyrightheading}[1]
	{\vspace*{-2.5cm}\smalllineskip{\flushleft
	{\footnotesize International Journal of Modern Physics D,
          #1}\\
	{\footnotesize $\eightcopyright$\, World Scientific Publishing
	 Company}\\
	 }}

\newcommand{\pub}[1]{{\begin{center}\footnotesize\smalllineskip
	Received #1\\
	\end{center}
	}}

\def\abstracts#1#2#3{{
	\centering{\begin{minipage}{4.5in}\baselineskip=10pt\footnotesize
	\parindent=0pt #1\par
	\parindent=15pt #2\par
	\parindent=15pt #3
	\end{minipage}}\par}}

\newcommand{\bibit}{\nineit}
\newcommand{\bibbf}{\ninebf}
\renewenvironment{thebibliography}[1]
        {\frenchspacing
	 \ninerm\baselineskip=11pt
         \begin{list}{\arabic{enumi}.}
        {\usecounter{enumi}\setlength{\parsep}{0pt}
	 \setlength{\leftmargin 12.7pt}{\rightmargin 0pt} 
         \setlength{\itemsep}{0pt} \settowidth
	{\labelwidth}{#1.}\sloppy}}{\end{list}}

\newcounter{itemlistc}
\newcounter{romanlistc}
\newcounter{alphlistc}
\newcounter{arabiclistc}

\newcommand{\fcaption}[1]{
        \refstepcounter{figure}
        \setbox\@tempboxa = \hbox{\footnotesize Fig.~\thefigure. #1}
        \ifdim \wd\@tempboxa > 5in
           {\begin{center}
        \parbox{5in}{\footnotesize\smalllineskip Fig.~\thefigure. #1}
            \end{center}}
        \else
             {\begin{center}
             {\footnotesize Fig.~\thefigure. #1}
              \end{center}}
        \fi}

\newcommand{\tcaption}[1]{
        \refstepcounter{table}
        \setbox\@tempboxa = \hbox{\footnotesize Table~\thetable. #1}
        \ifdim \wd\@tempboxa > 5in
           {\begin{center}
        \parbox{5in}{\footnotesize\smalllineskip Table~\thetable. #1}
            \end{center}}
        \else
             {\begin{center}
             {\footnotesize Table~\thetable. #1}
              \end{center}}
        \fi}

\def\@citex[#1]#2{\if@filesw\immediate\write\@auxout
	{\string\citation{#2}}\fi
\def\@citea{}\@cite{\@for\@citeb:=#2\do
	{\@citea\def\@citea{,}\@ifundefined
	{b@\@citeb}{{\bf ?}\@warning
	{Citation `\@citeb' on page \thepage \space undefined}}
	{\csname b@\@citeb\endcsname}}}{#1}}

\newif\if@cghi
\def\cite{\@cghitrue\@ifnextchar [{\@tempswatrue
	\@citex}{\@tempswafalse\@citex[]}}
\def\citelow{\@cghifalse\@ifnextchar [{\@tempswatrue
	\@citex}{\@tempswafalse\@citex[]}}
\def\@cite#1#2{{$\null^{#1}$\if@tempswa\typeout
	{IJCGA warning: optional citation argument
	ignored: `#2'} \fi}}

\def\pmb#1{\setbox0=\hbox{#1}
	\kern-.025em\copy0\kern-\wd0
	\kern.05em\copy0\kern-\wd0
	\kern-.025em\raise.0433em\box0}

\def\fnt#1#2{\footnotetext{\kern-.3em
	{$^{\mbox{\scriptsize #1}}$}{#2}}}

\def\fpage#1{\begingroup
\voffset=.3in
\thispagestyle{empty}\begin{table}[b]\centerline{\footnotesize #1}
	\end{table}\endgroup}

\def\runninghead#1#2{\pagestyle{myheadings}
\markboth{{\protect\footnotesize\it{\quad #1}}\hfill}
{\hfill{\protect\footnotesize\it{#2\quad}}}}
\font\tenrm=cmr10
\font\tenit=cmti10
\font\tenbf=cmbx10
\font\bfit=cmbxti10 at 10pt
\font\ninerm=cmr9
\font\nineit=cmti9
\font\ninebf=cmbx9
\font\eightrm=cmr8

\def\bh{${\Bbb R}^2\times{\Bbb S}^2\>$}
\def\qed{\hbox{${\vcenter{\vbox{  
   \hrule height 0.4pt\hbox{\vrule width 0.4pt height 6pt
   \kern5pt\vrule width 0.4pt}\hrule height 0.4pt}}}$}}
\begin{document}
\runninghead{Yu. P. Goncharov}
{Dirac Monopoles on Kerr Black Holes: Comparing Gauges}

\normalsize\textlineskip
\thispagestyle{empty}
\setcounter{page}{1}
\copyrightheading{Vol. 8, No. 1 (1999)}
\vspace*{0.88truein}
\fpage{1}
\centerline{\bf DIRAC MONOPOLES ON KERR BLACK HOLES:}
\vspace*{0.035truein}
\centerline{\bf  COMPARING GAUGES
\footnote{PACS Nos.: 04.20.Jb, 04.70.Dy, 14.80.Hv}}
\vspace*{0.37truein}
\centerline{\footnotesize YU. P. GONCHAROV}
\vspace*{0.015truein}
\centerline{\footnotesize\it Theoretical Group, Experimental Physics
Department, State
Technical University}
\baselineskip=10pt
\centerline{\footnotesize\it Sankt-Petersburg 195251, Russia}
\vspace*{0.225truein}
\pub{October 1998}
\vspace*{0.21truein}
\abstracts{
  We update our previous work on the description of
twisted configurations for complex massless scalar field on the
Kerr black holes as the sections of complex line bundles over
the Kerr black hole topology \bh. From physical point of view the appearance
of twisted configurations is linked with the natural presence of Dirac
monopoles that arise as connections in the above line bundles. We consider
their description in the gauge inequivalent to the one studied previously
and discuss a row of new features appearing in this gauge.
}{}{}
\vspace*{1pt}\textlineskip 
\section{Introductory Remarks} 
\vspace*{-0.5pt}
\noindent
  Recently in Ref.\cite{Gon97} we have described topologically inequivalent
configurations (TICs) of complex massless scalar field within the framework of
the Kerr black geometry. As was discussed in Ref.\cite{Gon97}, those TICs exist
owing to high nontriviality of the standard topology of the 4D black hole
spacetimes which is of the \bh-form. High nontriviality of the given topology
consists in the fact that over it there exist a huge (countable) number of
nontrivial real and complex vector bundles of any rank $N >1$ ( for complex
ones for $N=1$ too) and also the countable number of the so-called
Spin$^c$-structures. As a result,
there arises a nontrivial problem to take in theory into account the
possibilities tied with the existence of TICs. The latter ones can be treated
as both the sections and connections in the mentioned vector bundles. It is
clear that it is necessary to describe TICs in the form convenient to
physical applications. In its turn, the first step here is to describe TICs
of complex (massless) scalar field. For the case of
the Schwarzschild (SW) and Reissner-Nordstr\"{o}m (RN) metrics this was in main
features done in Refs.\cite{{Gon94},{Gon96}} and then was used when analysing
the contribution of twisted TICs of complex scalar field to the Hawking
radiation.\cite{Gon9697} The work of Ref.\cite{Gon97} extended the mentioned
description to the Kerr geometry case.

When deriving the results of Ref.\cite{Gon97} we were using some gauge for
the connections (Dirac monopoles) in complex line bundles over \bh-topology
needed to describe the above TICs (for more details see Ref.\cite{Gon97} and
below). In what follows we shall refer to this gauge as gauge I. There exists,
however, one more gauge (gauge II in what follows) inequivalent to gauge I,
which is also interesting
from physical point of view and which was not considered in Ref.\cite{Gon97}.
This gauge has been employed by us when building U($N$)-monopoles with $N>1$
in Refs.\cite{{Gon98a},{Gon98b}}. There is, therefore, some necessity to
compare both the gauges.
The present paper will just be devoted to the given question but we shall
restrict ouselves to the case of the U(1)(Dirac)-monopoles as the most
physically interesting one.

  Sec. 2 contains description of TICs and building $\rm U(1)$-monopoles in
the gauge II. Sec. 3 checks the Maxwell equations, Gauss theorem and
Lorentz condition for those monopoles
in the given gauge while Sec. 4 gives some estimates of their masses.
Sec. 5 considers one possible application of the results obtained to the
Hawking radiation from Kerr black holes and explores some differences arising
under this in both the gauges. Finally, Sec. 6 is devoted to the concluding
remarks.

  We use the ordinary set of the local Boyer-Lindquist coordinates
$t,r,\vartheta,\varphi$ covering the whole manifold \bh except for a set of
the zero measure. At this the surface $t$=const., $r$=const. is an oblate
ellipsoid with
topology ${\Bbb S}^2$ and the focal distance $a$ while $0\leq\vartheta<\pi,
0\leq\varphi<2\pi$. Under the circumstances we write
down the Kerr metric in the form

$$ ds^2=g_{\mu\nu}dx^\mu\otimes dx^\nu\equiv(1-2Mr/\Sigma)dt^2-
\frac{\Sigma}{\Delta}dr^2-\Sigma d\vartheta^2-
[(r^2+a^2)^2-\Delta a^2\sin^2\vartheta]
\frac{\sin^2\vartheta}{\Sigma}d\varphi^2 $$
$$+\frac{4Mra\sin^2\vartheta}{\Sigma}dtd\varphi     \eqno(1)$$

with $\Sigma=r^2+a^2\cos^2\vartheta$, $\Delta=r^2-2Mr+a^2$, $a=J/M$, where
$J, M$ are, respectively, a black hole mass and an angular moment.

  For inquiry let us adduce the components of metric in the cotangent
bundle of manifold \bh with the metric (1) (in tangent bundle), so long as
we shall need them in calculations below. These are

$$ g^{tt}=\frac{1}{\Sigma\Delta}[(r^2+a^2)^2-\Delta a^2\sin^2\vartheta],
g^{rr}=-\frac{\Delta}{\Sigma}, g^{\vartheta\vartheta}=-\frac{1}{\Sigma},
g^{\varphi\varphi}=-\frac{1}{\Delta\sin^2\vartheta}(1-2Mr/\Sigma),$$
$$ g^{t\varphi}=g^{\varphi t}=\frac{2Mra}{\Sigma\Delta}\>.  \eqno(2)$$

Besides we have $|g|=|\det(g_{\mu\nu})|=(\Sigma\sin\vartheta)^2$, $r_\pm=
M\pm\sqrt{M^2-a^2}$, so $r_+\leq r<\infty$, $0\leq\vartheta<\pi$,
$0\leq\varphi<2\pi$.

  Throughout the paper we employ the system of units with $\hbar=c=G=1$,
unless explicitly stated. Finally, we shall denote $L_2(F)$ the set of
the modulo square integrable complex functions on any manifold $F$
furnished with an integration measure.

\section{Description of TICs and Building Monopoles}

  As was discussed in Refs.\cite{{Gon94},{Gon96}}, TICs of
complex scalar field on the
4D black holes are conditioned by the availability of countable
number of complex line bundles over the \bh-topology
underlying the 4D black hole physics. Each TIC corresponds to the sections
of a complex line bundle $E$ while the latter can be characterized by its
Chern number $n\in{\Bbb{Z}}$ (the set of integers). TIC with $n=0$ can be
called {\it untwisted} one while the rest of the TICs with $n\not=0$ should be
referred to as {\it twisted}. Using the fact that all
the mentioned line bundles can be trivialized over the chart of
local coordinates $(t,r,\vartheta,\varphi)$ covering almost the whole manifold
\bh , one should try to obtain a suitable wave equation on the given chart for
massless TIC $\phi$ with the Chern number $n\in{\Bbb{Z}}$.

  When searching for the form of this wave equation we ought to follow
a number of reasonable principles that has been enumerated in
Ref.\cite{Gon97}.
But let us recall them here. It is obvious that the sought equation
should look as ${\cal D}^\mu{\cal D}_\mu\phi=0$, where ${\cal D}_\mu=
\partial_\mu-ieA_\mu$ is a covariant derivative on the sections of the
bundle $E$ with Chern number $n$, while ${\cal D}^\mu$ is a formal
adjoint to ${\cal D}_\mu$ regarding the usual scalar product in $L_2$(\bh)
induced by the metric (1). That is, the operator ${\cal D}^\mu$ acts on the
differential forms $b=b_\mu dx^\mu$ with coefficients in the bundle $E$ in
accordance with the rule

$${\cal D}^\mu(b)=-{1\over\sqrt{|g|}}\partial_\mu(g^{\mu\nu}
\sqrt{|g|}b_\nu)
+ig\overline{A_\mu}g^{\mu\nu}b_\nu\>,\eqno(3)$$
where the overbar signifies complex conjugation.

 Under this situation, the sought equation
should be derived by the standard procedure from the lagrangian

$${\cal L}=|g|^{1/2}g^{\mu\nu}\overline{{\cal D}_\mu\phi}
{\cal D}_\nu\phi\>,\eqno(4)$$

At $n=0$ the sought
equation should pass on to the known one for untwisted configuration of
Ref.\cite{Teu72}, while at $n\ne0,a=0$ it should be the wave equation for
twisted TICs on the Schwarzschild black hole.\cite{{Gon94},{Gon96}}
Finally, the sought equation at $n\ne0,a\ne0$ should keep the valuable
property of the variable separability which holds true for the $n=0,a\ne0$
case\cite{Teu72}, so long as it will be important when quantizing twisted
TICs.

To meet all the enumerated requirements we can employ the gauge freedom
in choice of the connection $A$ (vector-potential $A_\mu$) in the
bundle $E$. One such a choice was realised in Ref.\cite{Gon97} (gauge I).
In the present paper we shall realise another choice (gauge II) and
we shall take the following form for the sought equation

$${\cal D}^\mu{\cal D}_\mu\phi=\Box_n\phi=\Box\phi-
\frac{1}{\Sigma\sin^2\vartheta}\left[2in\cos\vartheta
\left(a\sin^2\vartheta\frac{\partial}{\partial t}+
\frac{\partial}{\partial\varphi}\right)-n^2\cos^2\vartheta\right]\phi=0
\eqno(5)$$

with the standard wave operator $\Box=|g|^{-1/2}\partial_\mu
(\sqrt{|g|}g^{\mu\nu}\partial_\nu)$ conforming to the metric (1).

Comparing the Eq. (5) to the wave equation derived from the lagrangian (4)

$$\Box\phi-\frac{ie}{\sqrt{|g|}}\partial_\mu(g^{\mu\nu}\sqrt{|g|}
A_\nu\phi)-(ie\overline{A_\mu}g^{\mu\nu}\partial_\nu+
e^2g^{\mu\nu}\overline{A_\mu}A_\nu)\phi=0\>,\eqno(6)$$

we obtain the (gauge) conditions

$$A_r=A_\vartheta=0\>,\eqno(7)$$
$$g^{tt}A_t+g^{t\varphi}A_\varphi=\frac{na\cos\vartheta}{e\Sigma}\>,
\eqno(8)$$
$$g^{\varphi t}A_t+g^{\varphi\varphi}A_\varphi=
\frac{n\cos\vartheta}{e\Sigma\sin^2\vartheta}\>.\eqno(9)$$

From here it follows

$$A_t=\frac{na\cos\vartheta}{e\Sigma},
A_\varphi=-\frac{n\cos\vartheta(r^2+a^2)}{e\Sigma}\>.\eqno(10)$$

Under the circumstances the connection in the bundle $E$ is $A=A_\mu dx^\mu=
A_t(r,\vartheta)dt+A_\varphi(r,\vartheta)d\varphi$ which
yields the curvature of the bundle $E$ as

$$F= dA=(\partial_tdt+\partial_rdr+
\partial_\vartheta d\vartheta+\partial_\varphi d\varphi)A=$$
$$-\partial_rA_tdt\wedge dr-\partial_\vartheta A_tdt\wedge d\vartheta
+\partial_rA_\varphi dr\wedge d\varphi+
\partial_\vartheta A_\varphi d\vartheta\wedge d\varphi\>. \eqno(11)$$

After this, if we integrate $F$ over any surface $t$=const., $r$=const,
we shall have

$$\int\limits_{S^2}\,F=
\int\limits_{S^2}\,\partial_\vartheta A_\varphi d\vartheta\wedge d\varphi=-
\frac{n}{e}\int\limits_{S^2}\,\Omega\sin\vartheta d\vartheta\wedge d\varphi=
-\frac{2\pi n}{e}\int\limits_0^\pi\,\Omega\sin\vartheta d\vartheta
\>\eqno(12)$$

with

$$\Omega=\frac{(r^2+a^2)(a^2\cos^2\vartheta-r^2)}{\Sigma^2} $$

and the direct evaluation gives the conventional Dirac charge quantization
condition

$$eq=4\pi n \eqno(13)$$

with magnetic charge $q=\int_{S^2}\,F$, so we can identify the coupling
constant $e$ with electric charge. As a consequence, the Dirac magnetic
$\rm U(1)$-monopoles naturally live on the Kerr black holes as connections in
complex line bundles and we get the whole family of them through the
relations (7)--(10), this family being parametrized by the Chern number
$n\in{\Bbb{Z}}$ of a complex line bundle over \bh.

Hence physically the appearance of TICs for complex scalar field should be
obliged to the natural presence of Dirac monopoles on black hole and due to
the interaction with them the given field splits into TICs.

Returning to the Eq. (5), we can use the ansatz
$\phi=(r^2+a^2)^{-1/2}e^{i\omega t}e^{-im\varphi}S(\vartheta)R(r)$ to obtain

$$\frac{d}{dr}\Delta\frac{d}{dr}\left(\frac{R}{\sqrt{r^2+a^2}}\right)+
\frac{(r^2+a^2)^2\omega^2-4Mmra\omega+m^2a^2}{\Delta}
\frac{R}{\sqrt{r^2+a^2}}=
-(\lambda+n^2)\frac{R}{\sqrt{r^2+a^2}}\>,   \eqno(14)$$

$$\frac{1}{\sin\vartheta}\frac{d}{d\vartheta}\sin\vartheta
\frac{dS}{d\vartheta}-\left(a^2\omega^2\sin^2\vartheta+2na\omega\cos\vartheta
+\frac{m^2+n^2-2mn\cos\vartheta}{\sin^2\vartheta}\right)S=
\lambda\,S\>.\eqno(15)$$

At $n=0$ the Eq. (15) coincides with the equation for the so-called
{\it oblate spheroidal functions} (see, e. g., Ref.\cite{Kom76}). After
replacing $x=\cos\vartheta,|x|\leq1$, the solution $S^l_m(a\omega,x)$ of (15)
has $l-|m|$ zeros in the interval (-1,1), so that always
$l\geq|m|, m\in{\Bbb{Z}}$. The corresponding eigenvalues
$\lambda=\lambda_{lm}(a\omega)$ increase with increasing $l$ and
$\lambda_{lm}\to\infty$ at $l\to\infty$.\cite{Kom76} But
$\lambda_{lm}(a\omega)$ cannot be expressed in an explicit form as a function
of $l,m,a\omega$, though $\lambda_{lm}(0)=-l(l+1)$.\cite{Kom76} On the other
hand, at $a=0$ the Eq. (15) coincides with the one for the functions
$P^l_{mn}(\cos\vartheta)$ of Refs.\cite{{Gon94},{Gon96},{Gon9697}} and
at this $l=|n|,|n|+1,...,|m|\leq l$. As a result, we conclude that at
$a\ne0,n\ne0$ the Eq. (15) has solutions $S^l_{mn}(a\omega,\cos\vartheta)$
conforming to the eigenvalue $\lambda_{nlm}(a\omega)$ with
$l=|n|,|n|+1,...,|m|\leq l$ and
$S^l_{mn}(0,\cos\vartheta)=P^l_{mn}(\cos\vartheta)$. It is evident that we
will have the orthogonality relation at $n$ fixed

$$\int\limits_0^\pi\,\overline{S^l_{mn}}(a\omega,\cos\vartheta)
S^{l^\prime}_{m^\prime n}(a\omega,\cos\vartheta)
\sin\vartheta d\vartheta={2\over2l+1}\delta_{ll^\prime}
\delta_{mm^\prime}\>,\eqno(16)$$

which does not explicitly depend on $a\omega$ and passes on to the relation
for $P^l_{mn}(\cos\vartheta)$ at $a=0$.\cite{{Gon94},{Gon9697}}

Under this situation the combinations
$Y_{nlm}(a\omega,\vartheta,\varphi)=
e^{-im\varphi}S^l_{mn}(a\omega,\cos\vartheta)$ should be called
the {\it monopole oblate spheroidal harmonics}, so that at $a=0$ we obtain
the {\it monopole spherical harmonics} $Y_{nlm}(\vartheta,\varphi)$
(concerning the latter ones see
Refs.\cite{{Gon94},{Gon96},{Gon9697},{Gon97N}} for more details).

It is clear that when quantizing twisted TIC with the Chern number $n$ we
can take the set of functions
$(r^2+a^2)^{-1/2}R_{nlm}^{a\omega}(r)e^{i\omega t}e^{-im\varphi}
S^l_{mn}(a\omega,\cos\vartheta)$ as a basis in $L_2$(\bh), where the functions
$R_{nlm}^{a\omega}(r)$ are the solutions of (14) conforming to
$\lambda=\lambda_{nlm}(a\omega)$, $|m|\leq l$, $l=|n|,|n|+1$,... At $n=0$ this
set coincides with the standard one used for quantizing the untwisted TIC (see,
e. g., Ref.\cite{Nov86}).

  If comparing gauge II under consideration with gauge I of Ref.\cite{Gon97}
one should note that in gauge I the component $A_\vartheta \ne 0$ and is
subject to the condition
$$\frac{i}{e}|g|^{-1/2}\partial_\vartheta(\sqrt{|g|}g^{\vartheta\vartheta}
A_\vartheta)+g^{\vartheta\vartheta}|A_\vartheta|^2=
-\frac{n^2}{e^2\Sigma\sin^2\vartheta}\left\{1-\frac{\cos^2\vartheta}
{\Sigma^2}\left[\left(r^2+a^2\right)^2-\Delta a^2\sin^2\vartheta\right]
\right\}\>.\eqno(7^\prime)$$
 As is seen from (7$^\prime$), generally speaking, $A_\vartheta\ne 0$ in
gauge I, if $n\ne 0$, so both the gauges are not equivalent to each other.
\vskip1cm
\section{Maxwell Equations, Gauss Theorem and Lorentz Condition}

We can define the Hodge star operator on 2-forms $F$ of any $k$-dimensional
(pseudo)riemannian manifold
$B$ provided with a (pseudo)riemannian metric $g_{\mu\nu}$ by the relation
(see, e. g., Ref.\cite{Bes87})
$$F\wedge\ast F=(g^{\mu\alpha}g^{\nu\beta}-g^{\mu\beta}g^{\nu\alpha})
F_{\mu\nu}F_{\alpha\beta}
\sqrt{|g|}\,dx^1\wedge dx^2\cdots\wedge dx^k \eqno(17)$$
in local coordinates $x^\mu$. In our case of the metric (1) this yields
$$\ast(dt\wedge dr)=\sqrt{|g|}(g^{\varphi t}g^{rr}dt\wedge d\vartheta+
g^{tt}g^{rr}d\vartheta\wedge d\varphi)\>,
\ast(dt\wedge d\vartheta)=-\sqrt{|g|}(g^{\varphi t}g^{\vartheta\vartheta}
dt\wedge dr + g^{tt}g^{\vartheta\vartheta}dr\wedge d\varphi)\>,$$
$$\ast(dt\wedge d\varphi)=\sqrt{|g|}(g^{tt}g^{\varphi\varphi}-
g^{\varphi t}g^{t\varphi})dr\wedge d\vartheta\>,
\ast(dr\wedge d\vartheta)=g^{rr}g^{\vartheta\vartheta}\sqrt{|g|}
dt\wedge d\varphi\>,$$
$$\ast(dr\wedge d\varphi)=-\sqrt{|g|}(g^{rr}g^{\varphi\varphi}
dt\wedge d\vartheta+g^{rr}g^{t\varphi}d\vartheta\wedge d\varphi)\>,
\ast(d\vartheta\wedge d\varphi)=\sqrt{|g|}
(g^{\vartheta\vartheta}g^{\varphi\varphi}dt\wedge dr
+g^{\vartheta\vartheta}g^{t\varphi}dr\wedge d\varphi)\>, \eqno(18)$$
so that $\ast^2=\ast\ast=-1$, as should be for the manifolds with lorentzian
signature.\cite{Bes87}

Then according to (11) we obtain

$$\ast F=(g^{t\varphi}g^{\vartheta\vartheta}\frac{\partial A_t}{\partial\vartheta}
+g^{\vartheta\vartheta}g^{\varphi\varphi}
\frac{\partial A_\varphi}{\partial \vartheta})
\sqrt{|g|}\,dt\wedge dr
-(g^{\varphi t}g^{rr}\frac{\partial A_t}{\partial r}+
g^{rr}g^{\varphi\varphi}\frac{\partial A_\varphi}{\partial r})
\sqrt{|g|}\,dt\wedge d\vartheta$$
$$+(g^{tt}g^{\vartheta\vartheta}\frac{\partial A_t}{\partial\vartheta}+
g^{\vartheta\vartheta}g^{t\varphi}
\frac{\partial A_\varphi}{\partial \vartheta})\sqrt{|g|}\,dr\wedge d\varphi-
(g^{tt}g^{rr}\frac{\partial A_t}{\partial r}+g^{rr}g^{t\varphi}
\frac{\partial A_\varphi}{\partial r})\sqrt{|g|}\,d\vartheta\wedge d\varphi
\>.\eqno(19)$$

It is clear that the Maxwell equations $dF=0$ are fulfilled for the form $F$
of (11) because $d^2=0$. On the other hand, for the Maxwell equations
$d\ast F=0$ to be fulfilled one needs in accordance with (19) to have

$$\frac{\partial}{\partial r}\left[\sqrt{|g|}\left(g^{rr}g^{\varphi t}
\frac{\partial A_t}{\partial r}+
g^{rr}g^{\varphi\varphi}\frac{\partial A_\varphi}{\partial r}\right)\right]
+\frac{\partial}{\partial \vartheta}\left[\sqrt{|g|}\left(
g^{t\varphi}g^{\vartheta\vartheta}\frac{\partial A_t}{\partial\vartheta}+
g^{\vartheta\vartheta}g^{\varphi\varphi}
\frac{\partial A_\varphi}{\partial \vartheta}\right)\right]=0\>,\eqno(20)$$

$$\frac{\partial}{\partial r}\left[\sqrt{|g|}\left(g^{tt}g^{rr}
\frac{\partial A_t}{\partial r}+g^{rr}g^{t\varphi}\frac{\partial A_\varphi}
{\partial r}\right)\right]
+\frac{\partial}{\partial \vartheta}\left[\sqrt{|g|}\left
(g^{tt}g^{\vartheta\vartheta}\frac{\partial A_t}{\partial\vartheta}+
g^{\vartheta\vartheta}g^{t\varphi}
\frac{\partial A_\varphi}{\partial \vartheta}\right)\right]=0\>.\eqno(21)$$

The direct calculation with using (2) and (10) shows that the equations
(20)--(21) are satisfied.
Also, as one can notice, from the expression for
$\ast F$ of (19) it follows

$$\int\limits_{S^2}\,\ast F=-\int\limits_{S^2}\,g^{rr}\left(g^{tt}
\frac{\partial A_t}{\partial r}+g^{t\varphi}\frac{\partial A_\varphi}
{\partial r}\right)\sqrt{|g|}d\vartheta\wedge d\varphi=
-\frac{4\pi anr}{e}\int\limits_{-1}^1\,\frac{x\,dx}{\Sigma^2}=0
\>,   \eqno(22)$$

where $x=\cos\vartheta$. Denoting the region inside the ellipsoid
$t$=const., $r$=const.
as $V$, we have in accordance with the Stokes theorem

$$\int\limits_V\,d\ast F=\int\limits_{\partial V=S^2}\,\ast F=0\>,\eqno(23)$$

i. e., the Gauss theorem is true for the monopoles under consideration.
This is enough for an external observer not to see any electric charge
of the Kerr black hole. Besides it should be emphasized that the total
(internal) magnetic charge $Q_m$ of black hole which
should be considered as the one summed up over all the monopoles remains
equal to zero because

$$Q_m=\frac{4\pi}{e}\sum\limits_{n\in{\Bbb{Z}}}\,n=0\>,\eqno(24)$$

so the external observer does not see any magnetic charge of the Kerr black
hole either though the monopoles are present on black hole in the sense
described above.

  Now defining the divergence of 1-form $A=A_\mu\,dx^\mu$ by the relation
(see, e.g., Ref.\cite{Bes87})
$${\rm div}(A)=\frac{1}{\sqrt{|g|}}\partial_\mu(\sqrt{|g|}g^{\mu\nu}
A_\nu)\>,$$
we shall for the given monopoles of (10) have
$${\rm div}(A)=\frac{1}{\sqrt{|g|}}\{\partial_t[\sqrt{|g|}
(g^{tt}A_t+g^{t\varphi}A_\varphi)]+\partial_\varphi[\sqrt{|g|}(g^{\varphi t}
A_t+ g^{\varphi\varphi}A_\varphi]\}=0\>,\eqno(25)$$
since nothing depends on $t$ and $\varphi$. As a result, the Lorentz
condition ${\rm div}(A)=0$ is fulfilled for the monopoles under discussion.

If comparing with gauge I of Ref.\cite{Gon97}, it should be noted that in it,
generally speaking, $d\ast F \ne 0$, but the Gauss theorem and Lorentz
condition are fulfilled as well. As a consequence, the equality $d\ast F=0$
holds true in the weak (integral) sense.\cite{Gon97}

\section{Monopole Masses}

We can now use the energy-momentum tensor $T_{\mu\nu}$ for
electromagnetic field

$$T_{\mu\nu}={1\over4\pi}(-F_{\mu\alpha} F_{\nu\beta} g^{\alpha\beta}+
{1\over4}F_{\beta\gamma}F_{\alpha\delta}g^{\alpha\beta}g^{\gamma\delta}
g_{\mu\nu}) \eqno(26)$$

to estimate masses of $\rm U(1)$-monopoles under consideration. In our case
$$T_{00}=\frac{1}{4\pi}[-g^{rr}F^2_{tr}-
g^{\vartheta\vartheta}F^2_{t\vartheta}+\frac{1}{4}g_{tt}(
g^{tt}g^{rr}F^2_{tr}+g^{tt}g^{\vartheta\vartheta}F^2_{t\vartheta}+
g^{rr}g^{\varphi\varphi}F^2_{r\varphi}+
g^{\vartheta\vartheta}g^{\varphi\varphi}F^2_{\vartheta\varphi})]
\>,    \eqno(27)$$

where $F_{tr}=-\partial_rA_t$, $F_{t\vartheta}=-\partial_\vartheta A_t$,
$F_{r\varphi}=\partial_rA_\varphi$,
$F_{\vartheta\varphi}=\partial_\vartheta A_\varphi$.

So long as we are in the asymptotically flat spacetime, we can calculate the
sought masses according to

$$m_{\rm mon}(n)=\int\limits_{t=const}\,T_{00}\sqrt{\gamma}\,
dr\wedge d\vartheta\wedge d\varphi\>,   \eqno(28)$$

where
$$\sqrt{\gamma}=\sqrt{\det(\gamma_{ij})}=\sqrt{\Sigma/\Delta}\sin\vartheta
\sqrt{(r^2+a^2)^2-\Delta a^2\sin^2\vartheta}   \eqno(29)$$

for the metric $d\sigma^2=\gamma_{ij}dx^i\otimes dx^j$ on the hypersurface
$t=const$, while $T_{00}$ is computed at the monopole with the Chern number
$n$.

Under this situation, it is not complicated
to check that $T_{00}\sqrt{\gamma}$ has the following
asymptotic behaviour at $r\to\infty$

$$T_{00}\sqrt{\gamma}\sim \frac{n^2\sin\vartheta}{16\pi e^2r^2}
  \>. \eqno(30)$$

As a result, we can estimate (in usual units)

$$m_{\rm mon}(n)\sim \frac{\hbar^2c^2}{G}
\frac{n^2}{16\pi e^2}4\pi\int\limits_{r_+}^{\infty}
\,\frac{dr}{r^2}=\frac{n^2}{4e^2r_+}\frac{\hbar^2c^2}{G} \>.\eqno(31)$$

Consequently, the conforming Compton wavelength
$$\lambda_{\rm{mon}}(n)=\frac{\hbar}{m_{\rm{mon}}(n)c}=\frac
{4e^2}{n^2}\frac{Gr_+}{\hbar c^3}\,,\eqno(32)$$
whereas the gravitational radius of black hole
$r_g=Gr_+/c^2\gg\lambda_{\rm{mon}}(n)$ for any $|n|\geq1$, so long as
$e^2/\hbar c\approx1/137$ with
$e=4.8\cdot10^{-10}\,{\rm cm^{3/2}\cdot g^{1/2}\cdot s^{-1}}$, so that
the given monopoles could reside as quantum objects in black holes.

  The monopoles with $n<0$ can be called {\it antimonopoles}. It is clear that
the masses of monopole and antimonopole with the same $|n|$ are equal. The
relation (31) along with the results of Refs.\cite{{Gon96},{Gon9697}} allow
us to draw the conclusion that for all three types of black holes
(Schwarzschild, Reissner-Nordstr\"{o}m and Kerr)
$m_{\rm mon}(n)\sim n^2/e^2r_+$, where $r_+$ is the characteristic size of
the horizon. The dependence of $m_{\rm mon}(n)$ from the Chern number $n$
($\sim n^2$) at first sight tells us that the availability of such objects
with the arbitrarily large masses on black holes should have a strong
influence on the own gravitational field of black hole. The study of the
contribution of twisted TICs, for example, to the Hawking
radiation\cite{Gon9697} shows, however, that really this contribution is
obliged only to the monopoles with $|n|$=1---10--15, i. e., one can conclude
that effectively the monopoles with big Chern numbers are absent or suppressed.
On the other hand, if we sum up over all the (anti)monopole masses, we shall
obtain the formally divergent result
$M_{\rm tot}=C\sum\limits_{n=1}^\infty\,n^2$ with some constant $C$. One
can, however, apply the standard $\zeta$-regularization method to rewrite
$M_{\rm tot}$ as $C\zeta(-2)$ with the Riemann zeta-function $\zeta(s)$. But
$\zeta(-2k)=0$, $k=1, 2,...$ (see, e. g., Ref.\cite{Abr64}) and, accordingly,
we obtain $M_{\rm tot}=0$. This result may point at a hidden mechanism of
cancellations among the monopole dynamical effects.

At the end of this section one can note that in gauge I the estimate of
monopole mass is practically the same but it requires excluding the unphysical
component $A_\vartheta$ with the help of gauge condition (7$^\prime$)
(see for more details Ref.\cite{Gon97}).

\section{Increase of Hawking Radiation from Kerr Black Hole}

When quantizing twisted TIC with the Chern number $n$ we shall
take the set of functions
$f_{nlm}^{a\omega}=(2\pi\omega)^{-1/2}(r^2+a^2)^{-1/2}R_{nlm}^{a\omega}(r)
e^{i\omega t}
Y_{nlm}(a\omega,\cos\vartheta)$ as a basis in $L_2$(\bh), where the functions
$R_{nlm}^{a\omega}(r)$ are the solutions of (14) conforming to
$\lambda=\lambda_{nlm}(a\omega)$, $|m|\leq l$, $l=|n|,|n|+1$,... and we shall
normalize the monopole oblate spheroidal harmonics
$Y_{nlm}(a\omega,\vartheta,\varphi)$ of Sec. 2 to 1, i. e., by the
condition
$$\int\limits_0^\pi\,\int\limits_0^{2\pi}
\overline{Y_{nlm}}(a\omega,\vartheta,\varphi)
Y_{nl^\prime m^\prime}(a\omega,\vartheta,\phi)
\sin\vartheta d\vartheta d\varphi=\delta_{ll^\prime}
\delta_{mm^\prime}\>.$$

After this we can evidently realize the procedure of quantizing
TIC with the Chern number $n$, as usual, by expanding this TIC in the modes
$f^{a\omega}_{nlm}$

$$\phi=\sum\limits_{l=|n|}^\infty\sum\limits_{|m|\leq l}
\int\limits_0^\infty\,d\omega
(a^-_{\omega nlm}f^{a\omega}_{nlm}+
b^+_{\omega nlm}\overline{f^{a\omega}_{nlm}})\,,$$
$$\phi^+=\sum\limits_{l=|n|}^\infty\sum\limits_{|m|\leq l}
\int\limits_0^\infty\,d\omega
(b^-_{\omega nlm}f^{a\omega}_{nlm}+
a^+_{\omega nlm}\overline{f^{a\omega}_{nlm}})\,,\eqno(33)$$

so that $a^\pm_{\omega nlm}$, $b^\pm_{\omega nlm}$ should be interpreted
as the corresponding
creation and annihilation operators for charged scalar particle in both the
gravitational field of black hole and the field of the conforming monopole
with the Chern number $n$ and we have the standard commutation relations
$$[a^-_i,\,a^+_j]=\delta_{ij},\>[b^-_i,\,b^+_j]=\delta_{ij}\>
\eqno(33^\prime)$$
and zero for all other commutators,
where $i =\{\omega nlm\}$ is a generalized index.

Let us now return to the Eq. (14). By replacing

$$r^*=r+\frac{M}{\sqrt{M^2-a^2}}\left(r_+\ln\frac{r-r_+}{r_+}-
r_-\ln\frac{r-r_-}{r_-}\right)   \eqno(34)$$

and by going to the dimensionless quantities $x=r^*/M$, $y=r/M$, $k=\omega M$,
Eq. (14) can be rewritten in the form

$$\frac{d^2\psi}{dx^2}+\left[k-\frac{m\alpha}{y^2(x)+\alpha^2}\right]^2
\psi=V_\lambda(x,k,\alpha)\psi\>,     \eqno(35)$$

where

$$V_\lambda(x,k,\alpha)=\left\{
-\frac{\lambda+n^2}{[y^2(x)+\alpha^2]^2}+\frac{\alpha^2}{[y^2(x)+\alpha^2]^3}
+\frac{2y(x)[y^2(x)-2\alpha^2]}{[y^2(x)+\alpha^2]^4}\right\}
[y^2(x)-2y(x)+\alpha^2]$$
$$-\frac{2km\alpha}{y^2(x)+\alpha^2}+
\frac{4km\alpha}{[y^2(x)+\alpha^2]^2}
  \eqno(36)$$

with $\alpha=a/M$, $\psi=\psi(x,k,\alpha)=R^*(Mx), R^*(r^*)=R[r(r^*)]$,
while $y(x)$ is the
one-to-one correspondence between $(-\infty,\infty)$ and
$(1+\sqrt{1-\alpha^2},\infty)$, so that
$$x=y+\frac{1}{\sqrt{1-\alpha^2}}\left(y_+\ln\frac{y-y_+}{y_+}-
y_-\ln\frac{y-y_-}{y_-}\right) \eqno(37)$$

with $y_\pm=1\pm\sqrt{1-\alpha^2}$, $0\leq\alpha\leq1$.
It is natural to search for the general solution of (35) in the class of
functions restricted on the whole $x$-axis in the linear combination form

$$\psi_{nlm}(x,k,\alpha)=C^1_{nlm}(k,\alpha)\psi^1_{nlm}(x,k,\alpha)+
C^2_{nlm}(k,\alpha)\psi^2_{nlm}(x,k,\alpha)\>,\eqno(38)$$
where the solutions $\psi^{1,2}_{nlm}(x,k,\alpha)$ pose a scattering problem
on the whole $x$-axis for the equation (35)

$$\psi_{nlm}^1(x,k,\alpha)\sim\cases{e^{i(k-m\alpha/2y_+)x}+
\ s_{12}(k,\alpha,n,l,m)e^{-i(k-m\alpha/2y_+)x},
&$x\rightarrow-\infty$,\cr
 s_{11}(k,\alpha,n,l,m)e^{ikx},&$x\rightarrow+\infty$,\cr} $$
$$\psi_{nlm}^2(x,k,\alpha)\sim\cases{s_{22}(k,\alpha,n,l,m)
e^{-i(k-m\alpha/2y_+)x},
&$x\rightarrow-\infty$,\cr
e^{-ikx}+\ s_{21}(k,\alpha,n,l,m)e^{ikx},&$x
\rightarrow+\infty$,\cr}\eqno(39)$$

with $S$-matrix
$$S(k,\alpha,n,l,m)=\pmatrix{s_{11}(k,\alpha,n,l,m)&s_{12}(k,\alpha,n,l,m)\cr
             s_{21}(k,\alpha,n,l,m)&s_{22}(k,\alpha,n,l,m)\cr}\>.\eqno(40)$$

  Having obtained these relations, one
can speak about the Hawking radiation process for any TIC of complex scalar
field on black holes. Actually, one should use the energy-momentum tensor
for TIC with the Chern number $n$ conforming to the lagrangian (4)

$$T_{\mu\nu}={\rm Re}[(\overline{{\cal D}_\mu\phi})({\cal D}_\nu\phi)
-{1\over2}g_{\mu\nu}g^{\alpha\beta}(\overline{{\cal D}_\alpha\phi})
({\cal D}_\beta\phi)]\>,\eqno(41)$$
or, in quantum form
$$T_{\mu\nu}={\rm Re}[(\overline{{\cal D}_\mu}\phi^+)({\cal D}_\nu\phi)
-{1\over2}g_{\mu\nu}g^{\alpha\beta}(\overline{{\cal D}_\alpha}\phi^+)
({\cal D}_\beta\phi)]\>.\eqno(42)$$

To get, according to the standard
receipt (see, e. g., Ref.\cite{Wit75}) but with
employing the monopole oblate spheroidal harmonics described in Sec. 2,
the luminosity $L(M,J,n)$ with respect to the Hawking radiation for T
HôÉhPM5=VXY€Ð
HôÉhPM5=VXY5
with
the Chern number $n$, we define a vacuum state $|0>$ by the conditions
$$a^-_i|0>=0, \> b^-_i|0>=0\>,\eqno(43)$$
and then

$$L(M,J,n)=\lim_{r\to\infty}\,\int\limits_{S^2}\,
<0|T_{tr}|0>d\sigma \eqno(44)$$

with the vacuum expectation value $<0|T_{tr}|0>$ and the surface element
$d\sigma=\sqrt{g_{\vartheta\vartheta}g_{\varphi\varphi}}d\vartheta\wedge
d\varphi$ tending to $r^2\sin\vartheta d\vartheta\wedge d\varphi$ when
$r\to+\infty$.
Under the circumstances with the help of (10), (33), (42) and (43)
we find at $r\to\infty$
$$<0|T_{tr}|0>\longrightarrow{\rm Re}\left[\sum_{l=|n|}^\infty\sum_{m=-l}^{l}
\int_0^\infty\>\overline{(\partial_rf^{a\omega}_{nlm}})(\partial_t
f^{a\omega}_{nlm})d\omega\right] \>.\eqno(45)$$
Further, using the explicit form of $f^{a\omega}_{nlm}$ and the mentioned
normalization of $Y_{nlm}(a\omega,\cos\vartheta)$ we shall obtain

$$L(M,J,n)=\lim_{r\to\infty}{\rm Re}\left[\frac{i}{2\pi}
\sum_{l=|n|}^\infty\sum_{m=-l}^{l}
\int_0^\infty\>\overline{(\partial_rR^{a\omega}_{nlm}})
R^{a\omega}_{nlm}d\omega\right] \>.\eqno(46)$$

Now we can pass on to the quantities $x$, $k$, $\alpha$ introduced early in
this section to get

$$L(M,J,n)=L(\alpha,n)=M^{-2}\lim_{x\to\infty}{\rm Re}\left[\frac{i}{2\pi}
\sum_{l=|n|}^\infty\sum_{m=-l}^{l}
\int_0^\infty\>\overline{(\partial_x\psi_{nlm}(x,k,\alpha)}
\psi_{nlm}(x,k,\alpha)dk\right] \>,\eqno(47)$$
where the relations
$$\frac{d}{dr}=\frac{r^2+a^2}{\Delta}\frac{d}{dr^*}=\frac{x^2+\alpha^2}
{(x^2-2x+\alpha^2)M}\frac{d}{dx}\eqno(48) $$
were used.

As a consequence, the choice of vacuum state can be defined by the choice of
the suitable linear combination of (38) for $\psi_{nlm}(x,k,\alpha)$. Under
this situation the Hawking radiation corresponds to the mentioned choice in
the form
$$C^1_{nlm}(k,\alpha)=\frac{1}{\sqrt{\exp(\frac{k-m\Omega_0}{TM})-1}},\>
C^2_{nlm}(k,\alpha)=0\>,\eqno(49)$$
which at last with using asymptotics (39) entails (in usual units)
$$L(\alpha,n)=
A\sum\limits_{l=|n|}^\infty\sum\limits_{|m|\leq l}
\int\limits_0^\infty\,\frac{|s_{11}(k,\alpha,n,l,m)|^2\,kdk}
{\exp(\frac{k-m\Omega_0}{TM})-1}\>, \eqno(50)$$

where the Kerr black hole
temperature $T=(y_+-y_-)/4\pi My_+^2$, $\Omega_0=\alpha/(y^2_++\alpha^2)$,
$A=\frac{1}{2\pi\hbar}\left(\frac{\hbar c^3}{GM}\right)^2\approx
0.273673\cdot10^{50}\,{\rm{erg\cdot s^{-1}}}\cdot M^{-2}$ ($M$ in g).

We can interpret $L(\alpha,n)$ as an additional contribution to the
Hawking radiation due to the additional charged scalar particles leaving
the black hole because of the interaction with monopoles and
the conforming radiation
can be called {\it the monopole Hawking radiation}. Under this situation,
for the total luminosity $L(\alpha)$ of black hole with respect
to the Hawking radiation concerning the complex scalar field to be obtained,
one should sum up over all $n$, i. e.
$$L(\alpha)=\sum\limits_{n\in{\Bbb{Z}}}\,L(\alpha,n)=L(\alpha,0)+
\sum\limits_{n=1}^\infty\,L(\alpha,-n)+\>
\sum\limits_{n=1}^\infty\,L(\alpha,n)\>,
\eqno(51)$$
where, generally speaking, $L(\alpha,-n)\ne L(\alpha,n)$ owing to the fact that
$\lambda_{nlm}(a\omega)\ne\lambda_{-nlm}(a\omega)$ for the eigenvalues
$\lambda$ from (15).

As a result, we can expect a marked increase of Hawking radiation from
Kerr black holes. But for to get an exact value of this increase one should
apply numerical methods, so long as the scattering problem (39) does not
admit any exact solution and is complicated enough for consideration ---
the potential $V_\lambda(x,k,\alpha)$ of (36) is given in an implicit form
and eigenvalues $\lambda=\lambda_{nlm}(a\omega)$ can be defined only
numerically which also constitutes the separate difficult task.
Therefore, one should put off obtainment of the numerical results for Kerr
black holes until the mentioned problems are overcome so we hope to obtain
those numerical results elsewhere.

 Returning to comparing the gauges I and II, it should be noted that the
equations (14) and (15) are somewhat different in gauge I as well as the
potential $V_\lambda(x,k,\alpha)$ of (36) (see for more details
Ref.\cite{Gon97}).
The same above problems for numerical computations in gauge I are yet unsolved
either and, therefore, for the aim of comparing two gauges I and II we
shall in the present section consider the case $a=0$ (as a result, $\alpha=0$),
i. e. the SW black hole case where the corresponding problems can
be solved. In both the gauges we shall instead of (37) have
$x=y+2\ln(0.5y-1)$, $-\infty<x<\infty$, $2\leq y<\infty$, $y(x)$ is
the one-to-one correspondence between $(-\infty,\infty)$ and $(2,\infty)$ and,
besides, $y(x)>0$ and monotonically increases at $x\in(2,\infty)$, so that
$y_+=2,y_-=0$. At $\alpha=0$ in both the gauges we have
$\lambda_{nlm}(0)=\lambda_{-nlm}(0)=-l(l+1), l=|n|,|n|+1,\cdots,|m|\leq l$
for the eigenvalues $\lambda$ and, therefore
$$L(0)=\sum\limits_{n\in{\Bbb{Z}}}\,L(0,n)=L(0,0)+
2\sum\limits_{n=1}^\infty\,L(0,n)\>,
\eqno(52)$$
because $L(0,n)=L(0,-n)$ here and
$$L(0,n)=A\sum\limits_{l=|n|}
^\infty(2l+1)c_{nlm}\>\eqno(53)$$
with
$$c_{nlm}=\int\limits_0^\infty\frac{|s_{11}(k,0,n,l,m)|^2\,k\,dk}
{e^{8\pi k}-1}\>.\eqno(54)$$
As for the potential $V_\lambda(x,k,0)$ of (36) then it is equal to
$$V_\lambda(x)=\left[1-\frac{2}{y(x)}\right]\,\left[
\frac{l(l+1)}{y^2(x)}+\frac{2}{y^3(x)}\right]\>\eqno(55)$$
in the gauge I and
$$V_\lambda(x)=\left[1-\frac{2}{y(x)}\right]\,\left[
\frac{l(l+1)-n^2}{y^2(x)}+\frac{2}{y^3(x)}\right]\>\eqno(56)$$
in the gauge II.

Under this situation, as the general theory shows
(see, e. g., Ref.\cite{CS77}),
in order $s_{11}(k,0,n,l,m)$ to exist for the scattering problem (39) for
the equation (35) at $\alpha=0$, it is enough to have
$$Q_\lambda=\int\limits_{-\infty}^{+\infty}|V_\lambda(x)|dx<\infty\>.
 \eqno(57)$$

As is not complicated to show,\cite{Gon9697} this condition is fulfilled
for both potentials (55)--(56) and we may apply the general
theory\cite{CS77} to express
$$s_{11}(k,0,n,l,m)=\frac{2ik}{2ik-J_\lambda(k)}\eqno(58)$$
with
$$J_\lambda(k)=\int\limits_{-\infty}^{+\infty}V_\lambda(x)e^{ikx}
f_\lambda^{-}(x,k)dx
=\int\limits_{-\infty}^{+\infty}V_\lambda(x)e^{-ikx}
f_\lambda^{+}(x,k)dx\>,\eqno(59)$$
where the so-called Jost functions $f_\lambda^\pm(x,k)$ satisfy the Volterra
integral equations
$$f_\lambda^{+}(x,k)=e^{ikx}-\int\limits_x^{\infty}\frac{\sin k(x-t)}{k}
V_\lambda(t)f_\lambda^{+}(t,k)dt\>,$$
$$f_\lambda^{-}(x,k)=e^{-ikx}+\int\limits_{-\infty}^x\frac{\sin k(x-t)}{k}
V_\lambda(t)f_\lambda^{-}(t,k)dt\>.\eqno(60)$$
  Besides, from (58)-(59) it follows that $s_{11}(0,0,n,l,m)=0$
if $J_\lambda(0)\not=0$. The latter condition can be checked only
numerically for potentials $V_\lambda(x)$ in question and this check certifies
that the condition is fulfilled, so, in what follows, we consider
$J_\lambda(0)\not=0$.
Finally, theory shows\cite{CS77} that at $k\rightarrow+\infty$
$$s_{11}(k,0,n,l,m)=\frac{1}{1+\frac{\beta}{k}+\frac{o(1)}{k}}$$
with $\beta=-\frac{1}{2i}\int_{-\infty}^{+\infty}V_\lambda(x)dx$, that is,
$|s_{11}(k,0,n,l,m)|^2\rightarrow1$ at $k\rightarrow{+\infty}$.

Now we can estimate the behaviour of $L(0,n)$ in both the gauges. For this
aim we represent the coefficients $c_{nlm}$ of (54) in the form (omitting
the integrand)
$$c_{nlm}=c_{nlm}^1+c_{nlm}^2=
\int\limits_0^{\ln l/2\pi}+\int\limits_{\ln l/2\pi}^\infty
\>,\eqno(61)$$
so that $L(0,n)$ of (53) is equal to $L_1(0,n)+L_2(0,n)$ respectively.
  From (58) it follows (the prime signifies differentiation with respect to
$k$)
$$s^{\prime}_{11}(0,0,n,l,m)=-\frac{2i}{J_\lambda(0)}\>.$$
Since in our case $s_{11}(0,0,n,l,m)=0$ (see above), then at
$0\leq k\leq \ln l/2\pi$
we can put $|s_{11}(k,0,n,l,m)|^2\sim|s^{\prime}_{11}(0,0,n,l,m)|^2k^2$.

\subsection{The gauge I}
One can rewrite (55) as $V_\lambda(x)=l(l+1)U_0(x)$ and considering
$f^-_\lambda(x,k)\sim e^{-ikx}$, we obtain (because $y(x)>0$, see above)
$$J_\lambda(0)\sim l(l+1)\int\limits_{-\infty}^\infty U_0(x)\,dx\geq
l(l+1)Q\>\eqno(62)$$
with
$$Q=\int\limits_{-\infty}^\infty\left[1-\frac{2}{y(x)}\right]\,
\frac{dx}{y^2(x)}>0\>,\eqno(63)$$
which entails
$$c_{nlm}^1\sim \frac{4}{[l(l+1)Q]^2}\int\limits_0^{\ln l/2\pi}
{k^3dk\over e^{8\pi k}-1}\leq \frac{4}{[l(l+1)Q]^2}\frac{1}{(8\pi)^4}
\int\limits_0^\infty\frac{k^3dk}{e^k-1}=$$
$$\frac{4}{[l(l+1)Q]^2(8\pi)^4}3!\,\zeta(4)
=\frac{1}{30\cdot8^3[l(l+1)Q]^2}\>,$$
where we used the formula of Ref.\cite{Abr64}
$$\int\limits_0^\infty{t^ndt\over e^t-1}=n!\,\zeta(n+1)$$
with the Riemann zeta function $\zeta(s)$, while $\zeta(4)=\pi^4/90$. Under
these conditions
$$L_1(0,n)=\frac{A}{2\pi}\sum\limits_{l=|n|}^\infty(2l+1)c^1_{nlm}\sim
\frac{A}{2\pi}\int\limits_{|n|}^\infty(2l+1)c^1_{nlm}\,dl=
\frac{A}{2\pi(n^2+|n|)30\cdot8^3Q^2}
\sim\frac{A}{2\pi n^2 30\cdot8^3Q^2}\>.\eqno(64)$$
  As for the coefficient $c^2_{nlm}$ then [since $|s_{11}(k,0,n,l,m)|^2\leq1$]
$$c^2_{nlm}=\int\limits_{\ln l/2\pi}^\infty\frac{|s_{11}(k,0,n,l,m)|^2\,k\,dk}
{e^{8\pi k}-1}\leq\int\limits_{\ln l/2\pi}^\infty k\,e^{-8\pi k}dk=
\frac{1}{16\pi^2l^4}\left(\ln l+\frac{1}{4\pi}\right)$$
and, as a consequence,
$$L_2(0,n)=\frac{A}{2\pi}\sum_{l=|n|}^\infty(2l+1)c_{nlm}^2\sim
\frac{A}{16\pi^3}
\int\limits_{|n|}^\infty\frac{\ln l dl}{l^3}=\frac{A}{64\pi^3}\,
\frac{2\ln n+1}{n^2}\>, \eqno(65)$$
and we can estimate that
$$L(0,n)\sim L_1(0,n)+L_2(0,n)\sim B\frac{\ln n}{n^2}\>\eqno(66)$$
with some constant $B$ not depending on $n$. As a result, the series over
$n$ in the right-hand side of (52) is convergent and total $L(0)<\infty$.
  Really, the numerical computation\cite{Gon9697} says that twisted
configurations on the SW black hole
gives the contribution of order 17\% to the total luminosity $L(0)$ of (52).

\subsection{The gauge II}
It is obvious that we shall here have for potential of (56)
$$J_\lambda(0)\sim [l(l+1)-n^2]\int\limits_{-\infty}^\infty U_0(x)\,dx\geq
[l(l+1)-n^2]Q\>\eqno(67)$$
which entails
$$c_{nlm}^1\sim
\frac{1}{30\cdot8^3[l(l+1)-n^2]^2Q^2}\>,$$
and further
$$L_1(0,n)=\frac{A}{2\pi}\sum\limits_{l=|n|}^\infty(2l+1)c^1_{nlm}\sim
\frac{A}{2\pi}\int\limits_{|n|}^\infty(2l+1)c^1_{nlm}\,dl=
\frac{A}{2\pi|n|30\cdot8^3Q^2}\>.\eqno(68)$$
At the same time, the above estimate of $c^2_{nlm}$ will remain true
so that now
$$L(0,n)\sim L_1(0,n)+L_2(0,n)\sim C\frac{1}{|n|}\>\eqno(69)$$
with some constant $C$ not depending on $n$. As a result, the series over
$n$ in the right-hand side of (52) is divergent and total $L(0)=\infty$.

\subsection{Comparison of gauges}
To compare the results obtained let us pass on to the flat $M\to 0,  J\to 0$
limit. Under the circumstances the wave equation in gauge I of Ref.\cite{Gon97}
will turn into the usual Klein-Gordon equation in the flat Minkowsky space
whose spatial metric is written in standard spherical coordinates $r,
\vartheta,\varphi$ while the wave equation (5) in gauge II will become
the usual Klein-Gordon equation in presence of Dirac magnetic monopole which
is usually employed in standard theory of the so-called flat magnetic monopoles
(see, e. g., Ref.\cite{Ros82}). There is, however, one subtlety which has
already been discussed in Ref.\cite{Gon96} but let us recall it here as
well.

Strictly speaking, if the words
" the flat space " signify the Minkowsky space then for it there exist
no monopole (abelian or not) solutions because the latter solutions are given
by the connections in nontrivial vector bundles over the spacetime\cite{At88}
while the trivial topology ${\Bbb{R}}^4$ of the Minkowsky space does not
admit any nontrivial vector bundles. In the traditional approach to monopoles
(see, e. g., the review of Ref.\cite{Ros82}), however, the trivial
topology ${\Bbb{R}}^4$ is tacitly replaced by the
${\Bbb{R}}\times {\Bbb{R}}^3\backslash \{0\}$-topology, where
 ${\Bbb{R}}^3\backslash \{0\}$
implies ${\Bbb{R}}^3$ with the origin (where the monopole is) discarded. But
the ${\Bbb{R}}^3\backslash \{0\}$ is homeomorphic (and even diffeomorphic)
to \bh and, as a result, ${\Bbb{R}}^4$-topology is
tacitly replaced by the \bh-topology, i. e., by the very same
which is under consideration in this paper. The metric on
\bh-topology in the conventional
approach is, however, the metric of (1) at $M = J = 0$.

When evaporating black hole it is natural to expect that the spacetime near
black hole becomes the genuine Minkowsky one, i. e., with topology
${\Bbb{R}}^4$ anf flat metric. Under this situation the monopoles should also
vanish, so long as their existence is conditioned by nontrivial spacetime
topology near black hole, though monopoles help the evaporation due to the
monopole Hawking radiation. All of that holds true in gauge I and
the contribution of the monopoles to the Hawking radiation proves to be
finite.
In gauge II the monopoles do not vanish after evaporation since the spacetime
topology do not change and the contribution of monopoles proves to be
infinite. The latter fact may be amended by the conjecture that in real
physical situation only the finite equal number monopoles and antimonopoles
can reside in black hole to remain its magnetic charge equal to zero. The
further fate of monopoles remains, however, unclear in gauge II.

\section{Concluding Remarks}
Within the present paper framewowk we did not touch upon the possibility
of solving the problem of statistical substantiation of the Kerr black hole
entropy with the help of U(1)-monopoles since it has been done in Ref. 6
(in gauge II) and it should be noted that considerations concerning this
do not practically differ in both the gauges (see, e. g., the SW and RN
black hole cases in Ref. 16 where the gauge I was used).
So the results of both the present paper and
Refs.\cite{{Gon97},{Gon94},{Gon96},{Gon9697},{Gon98a},{Gon98b},{Gon97N}}
show that the study of
${\rm U(N)}$-monopoles (with $N\geq1$) on
the 4D black holes is interesting and important task. Such an exploration 
might shed new light on black hole structure and the place of the black holes
in the generic scheme of quantum gravity. Really, as has been seen above,
when taking the mentioned monopoles into consideration there arise new
features in the Hawking radiation process such as the
{\it monopole Hawking radiation}. Besides there appears some {\it fine
structure} able to create statistical ensemble necessary for generating
black hole entropy.\cite{{Gon98b},{Gon97N}}
Also, we have seen that the conventional Dirac charge
quantization condition (13) is naturally fulfilled near black holes and this
can allow one to make the rather audacious assumption that {\it all or almost
all charged matter} was created from black holes, for example, in early stages
of the universe evolution which predetermined the electric charge quantization
all around the whole observed universe, though monopoles having been
responsible for it vanished after evaporation of black holes.

It is clear that one can pass on to the more realistic fields, in the first
turn, to electromagnetic and spinorial
ones. As has been discussed in
Refs.\cite{{Gon97},{Gon94},{Gon96},{Gon9697},{Gon98b}},
these fields also admit TICs on black holes
and, in consequence, one should expect, for instance, a certain modification
of the Hawking radiation process for them as well. The case of interest is
also to explore vacuum polarization for TICs near black holes where the new
marked contributions are expected to appear as well. One should try to
have constructed ${\rm U(N)}$-monopoles with $N\geq1$ on the generic 
Kerr-Newman black holes too. Finally, the above program can obviously be 
extended to include the important class of Tomimatsu-Sato metrics\cite{Tom72}
and their charged versions\cite{Ern73} which are natural deformations of 
Kerr and Kerr-Newman metrics and can be useful when analysing a number of 
problems related with the 4D black hole physics.

   All the mentioned problems require urgent study and we hope to continue
this exploration.

\nonumsection{Acknowledgement}
\noindent
    The work was supported in part by the Russian Foundation for
Basic Research (grant No. 98-02-18380-a) and by GRACENAS (grant No.
6-18-1997).

\nonumsection{References}

\end{document}